\documentclass[journal]{IEEEtran}
\usepackage{amsmath,amsfonts}
\usepackage{algorithmic}
\usepackage{algorithm}
\usepackage{array}
\usepackage[caption=false,font=normalsize,labelfont=sf,textfont=sf]{subfig}
\usepackage{textcomp}
\usepackage{stfloats}
\usepackage{url}
\usepackage{verbatim}
\usepackage{graphicx}
\usepackage{cite}
\usepackage{mathtools}
\usepackage{bm}
\usepackage{color}
\begin{document}
	
	\title{Information Rates of Channels with Additive White Cauchy Noise}

	\author{Shuqin Pang and  Wenyi Zhang,~\IEEEmembership{Senior Member, IEEE}
		% <-this % stops a space
		\thanks{The authors are with the  Department of Electronic Engineering and Information Science, University of Science and Technology of China, Hefei 230027,  China (e-mail: shuqinpa@mail.ustc.edu.cn;  wenyizha@ustc.edu.cn).
			
			The work has been supported by the National Natural Science Foundation of China through Grant 62231022.}}

	\maketitle
	
	\begin{abstract}
		Information transmission over discrete-time channels with memoryless additive noise obeying a Cauchy, rather than Gaussian, distribution, are studied. The channel input satisfies an average power constraint. Upper and lower bounds to such additive white Cauchy noise (AWCN) channel capacity are established. In the high input power regime, the gap between upper and lower bounds is within $0.5$ nats per channel use, and the lower bound can be achieved with Gaussian input. In the lower input power regime, the capacity can be asymptotically approached by employing antipodal input. It is shown that the AWCN decoder can be applied to additive white Gaussian noise (AWGN) channels with negligible rate loss, while the AWGN decoder when applied to AWCN channels cannot ensure reliable decoding. For the vector receiver case, it is shown that a linear combining receiver front end loses the channel combining gain, a phenomenon drastically different from AWGN vector channels.
	\end{abstract}
	

	\section{Introduction}
	\IEEEPARstart{I}{n} this paper, we consider a discrete-time channel with memoryless additive noise, subject to an average power constraint on its input.
	Unlike the normally considered additive white Gaussian noise (AWGN), we model the memoryless, i.e., temporally independent and identically distributed (i.i.d.), additive noise as a random variable following a zero-median Cauchy distribution. We call such channel an additive white Cauchy noise (AWCN) channel.

For practical communication scenarios, there are cases where the channel noise is typically observed  to be   impulsive, for example, underwater acoustic noises \cite{Machell89}, low-frequency atmosphere noises \cite{Shinde74tcom}, ignition noise \cite{Spaulding77tcom}, and certain electromagnetic noises and interference in man-made urban environments \cite{Middletontec77} \cite{Blackard93jsac}. These types of  impulsive noise are  often observed  with heavier tails than Gaussian noise \cite{Machell89}. The alpha-stable distributions have proven to be good models for impulsive noise (see, e.g., \cite{Laguna15}). The Cauchy distribution, as a well-known special case of alpha-stable distributions (with alpha parameter one) \cite{Nikias95}, is heavy-tailed, having neither mean nor convergent variance. Therefore the AWCN channel may serve as a reasonable reference model (see, e.g., \cite{Rappaporttcom66}) for understanding certain communication scenarios in which the noise possesses an impulsive nature.

	It has been known for long that communication receivers designed for AWGN channels generally introduce severe performance degradation in the impulsive noise case \cite{Spaulding77tcom}. Prior research on impulsive noise in communications has primarily focused on its signal processing aspect;  see, e.g., \cite{Nikias95, Aazhangtcom87, Kassam88, Vadali17, Zhou21icl} and references therein. Relatively few results on its information-theoretic aspect have been reported. In \cite{Kerpez93isit}, capacity bounds of a specific class of mixed-density impulsive-noise channels are derived. In \cite{Verdu90tit}, the capacity per unit cost of the AWCN channel is derived as an example. In \cite{Fahs14isit}, the capacity of the AWCN channel under a logarithmic constraint is derived. In \cite{Freitas17tit}, some capacity bounds for alpha-stable noise channels under an absolute moment constraint are obtained, excluding the AWCN case. In \cite{Wang11CL}, discretization and
Blahut-Arimoto algorithm have been employed to approximate the  capacity for channels with alpha-stable noise. In \cite{Larsson24twc},   achievable information rates for a massive multiple-input-multiple-output (MIMO) channel with Cauchy noise are studied and numerically evaluated.

	For a Cauchy noise, since its divergent variance can be interpreted as an infinite noise power, the usual notion of signal-to-noise ratio (SNR) of an AWCN channel is always zero for any finite average input power. Such a property appears to pose some conceptual and technical difficulties, especially from a traditional AWGN perspective. In this paper, via establishing upper and lower bounds to the AWCN channel capacity, we show that the information transmission performance can be characterized by an SNR-like parameter $\gamma = P/\lambda^2$, the ratio between the average input power and the square of the scale parameter of the Cauchy noise.
	
	We derive closed-form capacity upper and lower bounds for AWCN channels. In the regime of high input power, the lower bound based on entropy power inequality (EPI) and the upper bound based on a genie-aided argument both grow logarithmically in $\gamma$, with a gap not exceeding $0.5$ nats per channel use. In the regime of low input power, invoking a capacity-per-unit-cost analysis, the channel capacity is asymptotically proportional to $\gamma$ with a coefficient of $0.25$, and can be achieved by antipodal input.
	
	In addition, we investigate the robustness of the optimal decoder for AWGN, i.e., the nearest-neighbor decoding rule, in AWCN channels, showing that any non-zero information rate cannot be achieved. On the other hand, via computing the generalized mutual information (GMI) when applying the optimal decoder for AWCN to AWGN channels, we show that the rate loss is negligible.

 We further investigate the behavior of the vector receiver case, where the receiver has multiple branches each with an independent Cauchy noise added. In the regime of low input power, the asymptotic scaling law of channel capacity is identified via a capacity-per-unit-cost analysis. In the regime of high input power, however, the EPI degenerates and does not lead to a non-trivial lower bound. Instead, we study the performance of linear combining receiver front ends, and find that the resulting information rate loses the channel combining gain, in sharp contrast to AWGN vector channels, where the maximum ratio combining receiver leads to a power gain of the norm of the channel gain vector.
 	
	\section{Capacity Bounds}\label{sec:scalar}
	
	Consider a scalar real-valued channel model as
	\begin{equation}\label{eqn:AWCN}
		Y = X + Z,
	\end{equation}
	where the input $X$ satisfies an average power constraint $\mathbf{E}[X^2] = P$, and the noise $Z$ follows the (centered) Cauchy distribution with density
	\begin{equation}\label{eqn:cauchy-density}
		f(z) = \frac{\lambda}{\pi} \frac{1}{\lambda^2 + z^2},\quad -\infty < z < \infty. 
	\end{equation}
	
	The density (\ref{eqn:cauchy-density}) is a symmetric function with the origin, and $\lambda$ is called its scale parameter which controls the ``width'' of the density. Its median is zero, but it does not have finite moments of any order. In particular, its first-order moment (i.e., mean) does not exist and its second-order moment (i.e., power) is infinity. Among its properties (see, e.g., \cite{Verdu23entropy}), in this paper we mainly use the following basic ones: first, the differential entropy of $Z$ is $h(Z) = \log(4\pi\lambda)$; second, $Z$ can be represented as the ratio between two independent Gaussian random variables, as $U/V$, where $U \sim \mathcal{N}(0, \lambda^2)$ and $V \sim \mathcal{N}(0, 1)$ are independent; third, $Z$ is infinitely divisible, so that for any $k$ mutually independent (centered) Cauchy random variables $Z_i$ with scale parameter $\lambda_i$, $i = 1, \ldots, k$, the sum $\sum_{i = 1}^k a_i Z_i$ is still (centered) Cauchy, with scale parameter $\sum_{i = 1}^k |a_i| \lambda_i$.
	
	The capacity of the AWCN channel (\ref{eqn:AWCN}) is given by the general capacity formula for channels with continuous amplitudes% (see, e.g., \cite{Gallager68})
	\begin{equation}
		C=\sup_{X:\mathbf{E}[X^2]\leq P} I(X;Y).
		\label{eqn:channel capacity}
	\end{equation}
	Exact evaluation of the supremum in (\ref{eqn:channel capacity}) is analytically intractable. We hence turn to upper and lower bounding $C$.
	
	\subsection{Capacity Bounds}\label{subsec:capacity-bounds}
	
	We can obtain a capacity lower bound by letting $X$ be Gaussian and by applying the EPI; see, e.g.,  \cite[Sec. 17.7]{Cover06}:
	\begin{align}
		C \geq I(X; Y) &= h(Y) - h(Y|X)\nonumber\\
		&= h(X + Z) - h(Z)\nonumber\\
		&\geq h(X + Z') - h(Z)\nonumber\\
		&= \frac{1}{2} \log\left[2\pi e \left(P + \frac{8\pi \lambda^2}{e}\right)\right] - \log (4 \pi \lambda)\nonumber\\
		C_{\mathrm{lb}, \mathrm{epi}} &\coloneqq \frac{1}{2} \log\left(1 + \frac{e P}{8 \pi \lambda^2}\right),
	\end{align}
	where $Z'$ is a Gaussian random variable with mean zero and variance $8\pi \lambda^2/e$, which is obtained by matching its differential entropy with the differential entropy of $Z$, $\log (4\pi \lambda)$.
	
	Recalling that a Cauchy random variable can be represented as the ratio between two independent Gaussian random variables, we can rewrite the channel model (\ref{eqn:AWCN}) as
	\begin{equation}
		Y = X + \frac{U}{V},
	\end{equation}
	where $U \sim \mathcal{N}(0, \lambda^2)$ and $V \sim \mathcal{N}(0, 1)$ are independent. We then obtain a capacity upper bound by assuming that the decoder has access to $V$.
	This leads to the genie-aided capacity upper bound as
	\begin{equation}
		C \leq C_{\mathrm{ub}, \mathrm{genie}} \coloneqq \frac{1}{2} \mathbf{E}\left[\log\left(1+V^2\frac{P}{\lambda^2}\right)\right],
	\end{equation}
	where the expectation is with respect to $V\sim\mathcal{N}(0,1)$.
	
	The capacity per unit cost characterizes the capacity in the limit of vanishing average power $P$, and it also induces a capacity upper bound for all values of $P$. The capacity per unit cost of the AWCN channel (\ref{eqn:AWCN}) has been obtained in \cite[Sec. II, Example 4]{Verdu90tit} as $\frac{1}{4 \lambda^2}$, and hence we have the capacity upper bound
	\begin{equation}
		C \leq C_{\mathrm{ub}, \mathrm{cpuc}} \coloneqq \frac{P}{4 \lambda^2}.
		\label{eqn:cpuc}
	\end{equation}
	This capacity upper bound is asymptotically tight as $P \rightarrow 0$.
	
	\subsection{High-Power Regime}
	
	As $P \rightarrow \infty$, the EPI-based capacity lower bound behaves like
	\begin{equation}
		C_{\mathrm{lb}, \mathrm{epi}} = \frac{1}{2} \log \frac{P}{\lambda^2} + \frac{1}{2} \log \frac{e}{8\pi} + o(1),
		\label{eqn:upper bound 1}
	\end{equation}
	where (and throughout this paper) $o(1) \rightarrow 0$ as $P \rightarrow \infty$. On the other hand, the genie-aided capacity upper bound behaves like
	\begin{equation}
		C_{\mathrm{ub}, \mathrm{genie}} = \frac{1}{2} \log \frac{P}{\lambda^2} + \frac{1}{2} \mathbf{E}\left[\log V^2\right] + o(1).
		\label{eqn:upper bound 2}
	\end{equation}
	
	Inspecting the second terms  in (\ref{eqn:upper bound 1}) and 	(\ref{eqn:upper bound 2}), it can be calculated that $\frac{1}{2} \log \frac{e}{8\pi}\approx -1.1121$ and $\frac{1}{2} \mathbf{E}\left[\log V^2\right]\approx -0.6352$, both in nats. So we find that the gap between the capacity lower and upper bounds is within $0.5$ nats per channel use in the regime of high input power.
	
	\subsection{Low-Power Regime}
	
	In the regime of low input power, the EPI-based capacity lower bound is relatively loose. We may instead compute the channel mutual information under specific discrete input distributions. In particular, consider the following antipodal signaling:
	\begin{align}
		X = \begin{cases}
			-\sqrt{P} & \text{w.p. } 1/2,\\
			\sqrt{P}  & \text{w.p. } 1/2.
		\end{cases}
	\end{align}

	The resulting probability density function of output $Y$ is $p(y) = \frac{1}{2} f(y+\sqrt{P})+\frac{1}{2} f(y-\sqrt{P})$, which is a mixture of two Cauchy distributions with different location parameters ($-\sqrt{P}$ and $\sqrt{P}$ respectively). According to \cite[Ch. 6]{Nielsen2022handbook}, the differential entropy of $Y$ has been derived as
	\begin{equation}
		h(Y)=\log\frac{2\sqrt{\lambda^2+P}}{\lambda+\sqrt{\lambda^2+P}}+\log(4\pi\lambda).
	\end{equation}
	This leads to the channel mutual information as
	\begin{equation}
		I(X; Y) 
		=\log\frac{2\sqrt{\lambda^2+P}}{\lambda+\sqrt{\lambda^2+P}},
	\end{equation}
	which, as $P\to 0$, behaves like 
	\begin{equation}
		\frac{P}{4\lambda^2}+o\left(P\right),
		\label{eqn:l-p}
	\end{equation} 
	where $o(P) \rightarrow 0$ as $P \to 0$.
	
	Comparing (\ref{eqn:l-p}) and (\ref{eqn:cpuc}), we find that as $P\to 0$, the capacity scales with $P/\lambda^2$ with a coefficient of $0.25$, and this can be achieved by employing antipodal input.
	
	Summarizing the capacity bounds and their asymptotic behaviors, we thus conclude that the capacity of AWCN channels can be characterized by $\gamma = P/\lambda^2$, which may be interpreted as the SNR of AWCN channels.
	
\begin{figure}[!t]	
		\centering
		\includegraphics[width=0.44\textwidth]
		{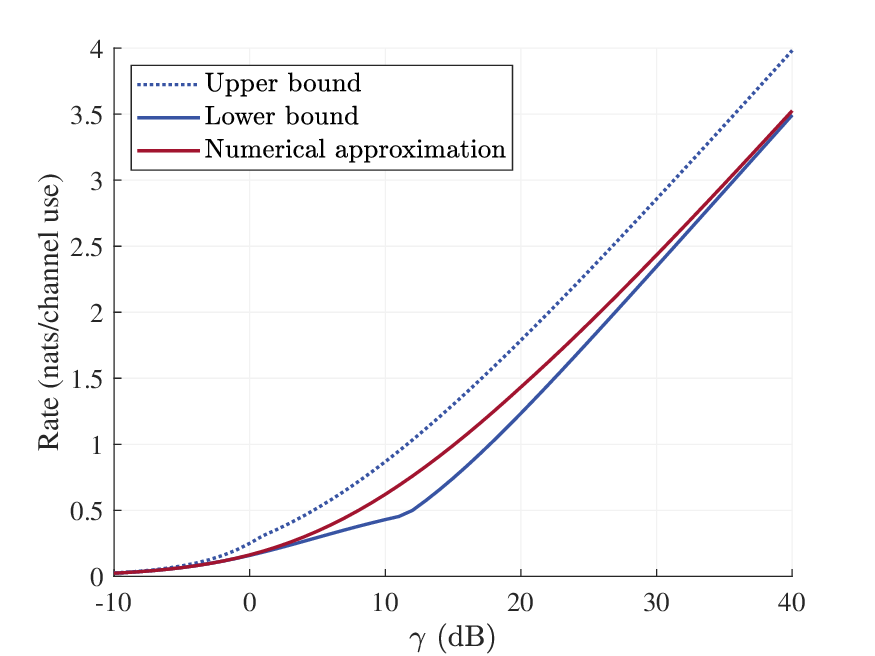}
		\caption{Upper and lower bounds of the AWCN capacity $C$, and $C$ numerically approximated using the Blahut-Arimoto algorithm for discretized input/output alphabets.}
		\label{fig:upper_lower_bound}
	\end{figure}

	For finite values of $\gamma$, Figure \ref{fig:upper_lower_bound} displays the capacity upper and lower bounds, and the capacity numerically approximated using the Blahut-Arimoto algorithm \cite{Blahut72tit} for discretized input/output alphabets.  We observe that the gap between the upper and lower bounds is always within $0.6$ nats per channel use, in the entire range of considered $\gamma$. As $\gamma$ grows large, the derived  channel lower bound gets close to the approximated channel capacity.
	
	\subsection{Robustness of Decoding}

	We proceed to examine the robustness of decoding. Specifically, we consider the information-theoretic performance when applying the optimal decoder for AWGN to AWCN channels, and the optimal decoder for AWCN to AWGN channels, respectively.
	
	Given a code rate $R$ (nats/channel use), the encoder chooses a message  $m$ from  the message set $\mathcal{M}=\{1,2,\cdots,\lceil e^{NR} \rceil\}$  uniformly  randomly, and maps it to a length-$N$ codeword $\bm x(m)$,  with elements $x_n(m)$, $n=1,2,\cdots,N$, for transmission. The resulting  channel output vector is $\bm y=(y_1,y_2,\cdots,y_N)$.
	For the AWCN channel, the optimal, i.e., maximum-likelihood (ML) decoding rule is
	\begin{align}
		\hat m&=\arg\max_{m\in\mathcal{M}} p(\bm y|\bm x(m))\nonumber\\
		&=\arg\min_{m\in\mathcal{M}}\sum_{n=1}^N\log\left[1+\frac{(y_n-x_n(m))^2}{\lambda^2}\right].
		\label{eqn:ML}
	\end{align}
	
	The ML decoding rule given by (\ref{eqn:ML}) has been utilized to develop the Viterbi decoder for AWCN channels in \cite{Kaiser01tsp}. The form of the AWCN ML decoding rule can be interpreted as a bent nearest-neighbor decoding rule. If $|y_n-x_n(m)|$ is small, then $\log\left[1+\left(\frac{y_n-x_n(m)}{\lambda}\right)^2\right]\approx (1/\lambda^2)\cdot(y_n-x_n(m))^2$, which is proportional to the Euclidean distance between $y_n$ and $x_n(m)$. However, if $|y_n-x_n(m)|$ gets larger, the logarithmic operation becomes effective, bending the Euclidean distance toward its logarithm. Intuitively, such a bending effect immunizes the decoder from the disturbance of exceedingly large noise samples, which occur with relatively high frequency for Cauchy distributions. 
	
	Without bending, the original nearest-neighbor decoding rule is ML for AWGN channels, but it performs poorly for AWCN channels. In \cite{Lapidoth96tit}, it has been shown that for an additive-noise channel with noise variance $\sigma^2$ and average input power constraint $P$, using Gaussian inputs and the nearest-neighbor decoding rule exactly achieves the information rate of $\frac{1}{2}\log\left(1+P/\sigma^2\right)$, as if the channel is AWGN. Because Cauchy distributions have divergent variances, i.e., $\sigma^2\to \infty$, the corresponding achievable rate thus vanishes.
	
	We can further strengthen the result by showing that any non-zero information rate cannot be achieved by applying the nearest-neighbor decoding rule for AWCN channels. It suffices to analyze the behavior of a codebook with only two codewords. Consider two length-$N$ codewords, $\bm x(1)$ and $\bm x(2)$. Assume that $\bm x(1)$ is transmitted and $\bm y$ is received. The nearest-neighbor decoding rule forms the two statistics:
	\begin{align}
		S_1&=\sum_{n=1}^N[y_n-x_n(1)]^2=\sum_{n=1}^Nz_n^2;\nonumber\\
		S_2&=\sum_{n=1}^N[y_n-x_n(2)]^2
		=\sum_{n=1}^N[x_n(1)-x_n(2)+z_n]^2,
	\end{align}
	where $z_n$, $n=1,2,\cdots, N$, are i.i.d. Cauchy noise samples. A decoding error occurs if $S_1 > S_2$,\footnote{When $S_1 = S_2$ a decoding error occurs with probability $1/2$, but this event occurs with zero probability and hence can be omitted in the subsequent analysis.} i.e.,
	\begin{equation}
		%&\sum_{n=1}^Nz_n^2 > \sum_{n=1}^N[x_n(1)-x_n(2)+z_n]^2, \ \mbox{or},\nonumber\\
		\frac{1}{N}\sum_{n=1}^N[x_n(2)-x_n(1)]z_n
		> \frac{1}{2N}\sum_{n=1}^N[x_n(2)-x_n(1)]^2.
		\label{eqn:e-o-condition}
	\end{equation}
	The left hand side of (\ref{eqn:e-o-condition}) is a Cauchy random variable with parameter $\frac{\lambda}{N}\sum_{n=1}^N|x_n(2)-x_n(1)|$, and therefore the decoding error probability is essentially the tail probability of this Cauchy random variable exceeding $\frac{1}{2N}\sum_{n=1}^N[x_n(2)-x_n(1)]^2$, which can be evaluated as 
	\begin{equation}\label{eqn:Pe}
		\varepsilon=\frac{1}{2}-\frac{1}{\pi}\arctan\left[\frac{\frac{1}{2N}\sum_{n=1}^N[x_n(2)-x_n(1)]^2}{\frac{\lambda}{N}\sum_{n=1}^N|x_n(2)-x_n(1)|}\right],
	\end{equation}
	by utilizing the cumulative distribution function (CDF) of a (centered) Cauchy random variable with scale parameter $\lambda$, $\frac{1}{\pi} \arctan \frac{z}{\lambda} + \frac{1}{2}$.
	
	As $N\to\infty$, $\varepsilon$ vanishes if and only if the argument of the $\arctan$ function in (\ref{eqn:Pe}) grows unbounded. Considering random coding, each symbol $x_n(i)$ is drawn independently from a distribution $p(x)$ satisfying the average power constraint $P$. Therefore from the law of large numbers, the two-codeword decoding error probability converges as
	\begin{equation}
		\varepsilon\to\frac{1}{2}-\frac{1}{\pi}\arctan\left[\frac{1}{2\lambda}\frac{\mathbf{E}[(X_2-X_1)^2]}{\mathbf{E}[|X_2-X_1|]}\right], \ \mbox{w.p. 1},
	\end{equation}
	where $X_1$ and $X_2$ are two independent random variables following $p(x)$. So we conclude that even the two-codeword decoding error probability does not vanish. For example, if $x_n(i)\sim\mathcal{N}(0,P)$, $i=1,2$, $n=1,2\cdots, N$, then
	$\varepsilon\to\frac{1}{2}-\frac{1}{\pi}\arctan\left[\frac{1}{2}\sqrt{\frac{\pi P}{\lambda^2}}\right]$, w.p. 1; if $x_n(i)$ obeys the antipodal signaling, then
	$\varepsilon\to\frac{1}{2}-\frac{1}{\pi}\arctan\left[\sqrt{\frac{P}{\lambda^2}}\right]$, w.p. 1.
	
	The preceding argument hence illustrates that the optimal decoder for AWGN is not robust in AWCN channels.
	
	On the other way around, we proceed to examine the performance of the optimal decoder for AWCN (\ref{eqn:ML}) for AWGN channels.
	Because the decoding metric in (\ref{eqn:ML}) is mismatched to the channel, we adopt the information-theoretic tool of GMI for measuring the achievable information rate. In fact, GMI is a lower bound of the mismatch capacity, corresponding to codebooks consisting of i.i.d. random variables \cite{kaplan93aeu}. We evaluate GMI using the following general expression \cite{ganti00it}: 
	\begin{equation}
		\label{eqn:dual form}
		I_\mathrm{GMI} = \sup_{\theta < 0}\mathbf{E}\left[\log\frac{e^{\theta d(X,Y)}}{\mathbf{E}_{(X)}\left[e^{\theta d(X,Y)}\right]}\right],
	\end{equation}
	where $d$ is the decoding metric adopted and $\mathbf{E}_{(X)}$ denotes the expectation taken with respect to $X$ only.
	
	Under the Gaussian input $X\sim\mathcal{N}(0,P)$ and the AWCN decoding metric $d(x,y)=\log\left[1+\left(\frac{y-x}{\lambda}\right)^2\right]$, we can rewrite the GMI (\ref{eqn:dual form}) as 
	\begin{align}
		\label{eqn:d-gmi}
		I_\mathrm{GMI} = \sup_{\theta < 0}\Bigg\{&\theta\mathbf{E}[\log(\lambda^2+(Y-X)^2)]\nonumber\\
		&-\mathbf{E}\left[\log\mathbf{E}_{(X)}\left[(\lambda^2+(Y-X)^2)^{\theta}\right]\right]\Bigg\},
	\end{align}
	which can be evaluated numerically.
	
\begin{figure}[!t]	
		\centering
		\includegraphics[width=0.44\textwidth]
		{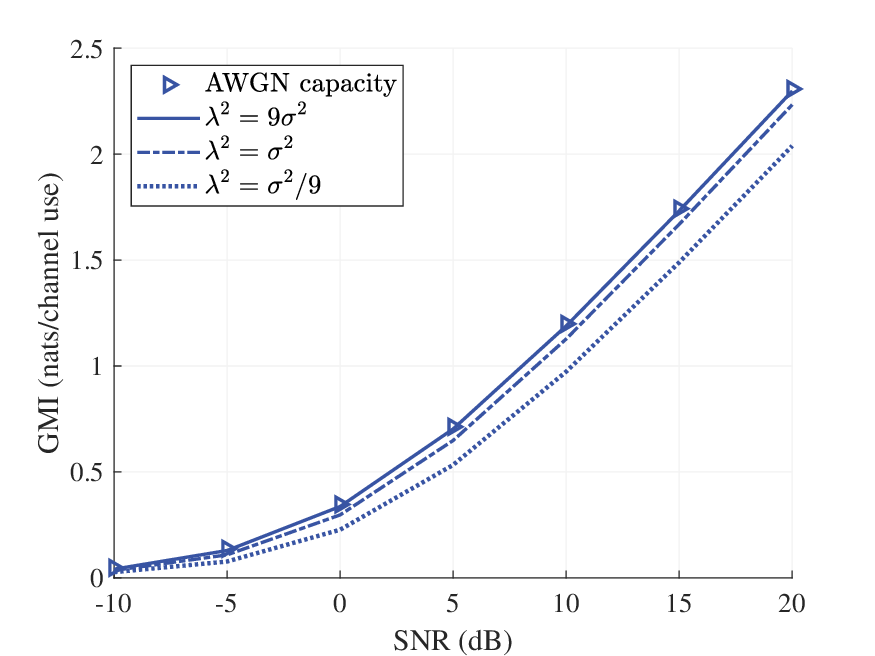}
		\caption{GMI under AWCN ML decoder in AWGN channels, for different choices of $\lambda^2$.}
		\label{fig:GMI_lambda}
	\end{figure}

	Figure \ref{fig:GMI_lambda} displays the GMI under the AWCN decoding metric for several different choices of $\lambda^2$ ($\sigma^2/9$, $\sigma^2$ and $9\sigma^2$) in AWGN channels. Note that since the channel model here is AWGN, the scale parameter $\lambda$ does not correspond to any physical aspect of the model and can be freely chosen as part of the decoder configuration. We observe that for $\lambda^2 = 9\sigma^2$, the GMI is very close to the AWGN capacity $\frac{1}{2}\log\left(1+P/\sigma^2\right)$, demonstrating the robustness of the AWCN ML decoder in AWGN channels. In fact, as discussed regarding the AWCN decoding metric (\ref{eqn:ML}), when $\lambda \gg \sigma$, the value of $(y-x)^2/\lambda^2$ is typically small, and thus the logarithmic bending effect of the AWCN decoding metric is mild.
	
	\section{Discussion on Vector Case}\label{sec:vector}
	
	In this section, we consider a real-valued channel model with a scalar input $X$ and a $k$-dimensional vector output $\underline{Y}$ as
	\begin{equation}
		\underline{Y} = \underline{h} X + \underline{Z},
	\end{equation}
	where the input $X$ satisfies an average power constraint $\mathbf{E}[X^2] = P$, different branches of $\underline{Y}$ have different channel gains given by $\underline{h}$, and the noise $\underline{Z}$ comprises of $k$ i.i.d. Cauchy random variables with scale parameter $\lambda$. We assume that the decoder knows $\underline{h}$.
	
The lower bounding technique utilizing the EPI is no longer applicable here. This is because the probability distribution of $\underline{h} X$ is degenerated in the $k$-dimensional Euclidean space, and hence the differential entropy of $\underline{h} X$ is minus infinity, leading to a trivial capacity lower bound of zero.
 
	When applying a linear combining vector $\underline{\beta}$ to process the output vector $\underline{Y}$, we obtain:
	\begin{equation}
		\underline{\beta}^T \underline{Y} = \underline{\beta}^T \underline{h} X + \underline{\beta}^T \underline{Z},
	\end{equation}
	which may be rewritten as:
	\begin{equation}\label{eqn:channel-linear-combined}
		\tilde{Y} = \tilde{h} X + \tilde{Z}.
	\end{equation}
	According to the properties of Cauchy distribution, $\tilde{Z}$ is still Cauchy, with parameter $\sum_{i = 1}^k |\beta_i| \lambda$. Without loss of generality, we may let all the elements of $\underline{h}$ be non-negative. Consequently, all the elements of $\underline{\beta}$ should be non-negative as well, and we further normalize $\underline{\beta}$ to satisfy $\sum_{i = 1}^k \beta_i = 1$. This way, $\tilde{Z}$ is Cauchy with parameter $\lambda$.

Invoking the EPI-based capacity lower bound in Section \ref{subsec:capacity-bounds}, we obtain a lower bound of the rate achieved by a linear combining receiver front end, as
\begin{eqnarray}
    C \geq \frac{1}{2} \log\left(1 + \frac{e \tilde{h}^2 P}{8\pi\lambda^2}\right).
\end{eqnarray}

	Optimizing the value of $\tilde{h}^2$ over non-negative $\underline{\beta}$ that satisfies $\sum_{i = 1}^k \beta_i = 1$, we can find that the maximum value of $\tilde{h}^2$ is $h_{\mathrm{max}}^2$, where $h_{\mathrm{max}}$ is the maximum value in $\{h_1,\cdots,h_{k}\}$. Therefore we obtain a capacity lower bound as
 \begin{eqnarray}
    C \geq \frac{1}{2} \log\left(1 + \frac{e h_{\mathrm{max}}^2 P}{8\pi\lambda^2}\right).
\end{eqnarray}
 From this, we notice that the linear combining receiver front end loses the channel combining gain, and only retains the selection gain, in power. Such a behavior is in sharp contrast to AWGN vector channels, where if one employs a maximum ratio combining receiver (i.e., $\underline{\beta} = \underline{h}$) then a power gain of $\|\underline{h}\|^2$ is realized.

Regarding capacity upper bound, similar to the scalar case in Section \ref{subsec:capacity-bounds}, by representing each element of $\underline{Z}$, $Z_i$, $i = 1, \ldots, N$, as ratio $U_i/V_i$, with $U_i \sim \mathcal{N}(0, \lambda^2)$ and $V_i \sim \mathcal{N}(0, 1)$ being independent, and by assuming that $\underline{V}$ is known at the decoder, we can obtain a genie-aided capacity upper bound, as
	\begin{align}
		C &\leq \max_{p(x): \mathbf{E}[X^2] = P} I(X; \underline{Y} | \underline{V})\nonumber\\
		&= \frac{1}{2} \mathbf{E}\left[\log \left(1 + \frac{\sum_{i = 1}^k h_i^2 V_i^2 P}{\lambda^2}\right)\right],
		\label{eqn:c_u_v}
	\end{align}
	where the expectation is taken with respect to $\underline{V}$. Expanding (\ref{eqn:c_u_v}) for sufficiently large $P$, we obtain the following asymptotic behavior:
	\begin{equation}
		C \leq \frac{1}{2} \log \frac{P}{\lambda^2} + \frac{1}{2} \mathbf{E}\left[\log \sum_{i = 1}^k h_i^2 V_i^2\right] + o(1),
	\end{equation}
%	where $o(1) \rightarrow 0$ as $P \rightarrow \infty$.
	where the second term corresponds to a power gain. 
	Using Jensen's inequality, we can further relax this term as
	\begin{equation}
		\frac{1}{2} \mathbf{E}\left[\log \sum_{i = 1}^k h_i^2 V_i^2\right] \leq \frac{1}{2} \log \|\underline{h}\|^2,
	\end{equation}
 indicating that the power gain is smaller than $\|\underline{h}\|^2$ in the high-power regime.

 So from the preceding analysis we can only conclude that the power gain lies between $h_{\mathrm{max}}^2$ and $\|\underline{h}\|^2$ in the high-power regime, but its exact value is still unknown.
	
	Besides, analogous to \cite[Sec. II, Example 4]{Verdu90tit}, the capacity per unit cost for the vector case can be shown to be $\frac{\|\underline{h}\|^2}{4\lambda^2}$, and this leads to the asymptotic capacity $\frac{\|\underline{h}\|^2 P}{4\lambda^2} + o(P)$ in the lower-power regime.

	\section{Conclusion}
	
	The main takeaway of this paper is that by defining the ``SNR'' as $\gamma = P/\lambda^2$, an AWCN channel behaves essentially similar to an AWGN channel, in the following senses: for large $\gamma$ the capacity grows logarithmically in $\gamma$, for small $\gamma$ the capacity is linear in $\gamma$. Such behaviors can only be realized by employing decoders matched to AWCN channels, and rate losses when one uses AWGN decoders are substantial. Furthermore, the exact power gain in the vector case and whether it can be achieved by a linear combining receiver front end still remain unknown problems.

	%\vfill
	
\end{document}